# STOKES FLOW OF INCOMPRESSIBLE LIQUID THROUGH A CONICAL DIFFUSER WITH PARTIAL SLIP BOUNDARY CONDITION

Peter Lebedev-Stepanov

Shubnikov Institute of Crystallography, Kurchatov Complex of Crystallography and Photonics, Leninskii prospekt 59, Moscow 119333, Russia

E-mail: lebstep.p@crys.ras.ru

**ABSTRACT**

An alternative form of the general solution of the linearized stationary axisymmetric Navier–Stokes equations for an incompressible fluid in spherical coordinates is obtained by the vector potential method. A previously published solution to this problem, dating back to the paper by Sampson, is given in terms of a stream function, which leads to formulas that are difficult to apply in practice. The presented form of solution is applied to the problem of liquid flowing through a conical diffuser under a partial slip boundary condition for a certain slip length $\lambda$. Recurrent relations are obtained that allow us to determine the velocity, pressure and stream function. The solution is analyzed in the first order of decomposition with respect to a small dimensionless parameter $\frac{\lambda}{r}$. It is shown that the sliding of the liquid over the surface of the cone leads to a vorticity of the flow. At zero slip length, we obtain the well-known solution to the problem of a diffuser with no-slip boundary condition corresponding to strictly radial streamlines.

## I. INTRODUCTION

In recent decades, investigations of the interaction of a liquid flow with a solid surface with partial slip boundary conditions have been rapidly developing [1-5]. This is due to the revolutionary development of technologies for creating functionalized surfaces, including hydrophobic and superhydrophobic coatings. The velocity of the aqueous flow in contact with such a solid coating is non-zero [3-8].

There are studies of the nature of this phenomenon at the atomic and molecular levels, as well as calculations of flow near surfaces of different shapes: flat, cylindrical, and spherical or their combinations [1,2,8,9]. Small-sized surfaces and slow fluid flow are usually considered that corresponds to small Reynolds number. For example, the sedimentation of a small spherical particle has been theoretically and experimentally considered. The drag force acting on a solid sphere moving through a fluid under a partial slip condition is described by generalization of the well-known Stokes' Law previously obtained for the non-slip condition [1, 10].

So-called Stokes equations are a linearized form of the stationary Navier–Stokes equations for incompressible liquids [11-13]:

$$\eta\overrightarrow{\nabla^2}\mathbf{V} = \nabla p, \quad (\nabla \cdot \mathbf{V}) = 0, \tag{1.1}$$

where $\mathbf{V}$ and $\eta$ are the velocity and the dynamic viscosity of a liquid, respectively, $p$ is the pressure. Vector Laplacian is determined by [14]

$$\overrightarrow{\nabla^2}\mathbf{V} = \nabla(\nabla \cdot \mathbf{V}) - \nabla \times [\nabla \times \mathbf{V}]. \tag{1.2}$$

Equations (1.1)-(1.2) in a spherical coordinate system [Fig. 1] for an axisymmetric problem can be rewritten as [12]:

$$\nabla^2 V_r - \frac{2V_r}{r^2} - \frac{2}{r^2 \sin\theta}\frac{\partial}{\partial \theta}(V_\theta \sin\theta) = \eta^{-1}\frac{\partial}{\partial r}p, \tag{1.3}$$

$$\nabla^2 V_\theta - \frac{V_\theta}{r^2 \sin^2\theta} + \frac{2}{r^2}\frac{\partial}{\partial \theta}V_r = \eta^{-1}\frac{1}{r}\frac{\partial}{\partial \theta}p, \tag{1.4}$$

$$\frac{1}{r}\frac{\partial}{\partial r}(r^2 V_r) + \frac{1}{\sin\theta}\frac{\partial}{\partial \theta}(V_\theta \sin\theta) = 0, \tag{1.5}$$

where the nonzero components of velocity $V_r(r,\theta)$, $V_\theta(r,\theta)$, and pressure $p$ are independent of the azimuthal angle $\varphi$.

A cone is a geometric figure that traditionally attracts attention because of its importance in applications and the relative simplicity of the mathematical formulation of the boundary problem.

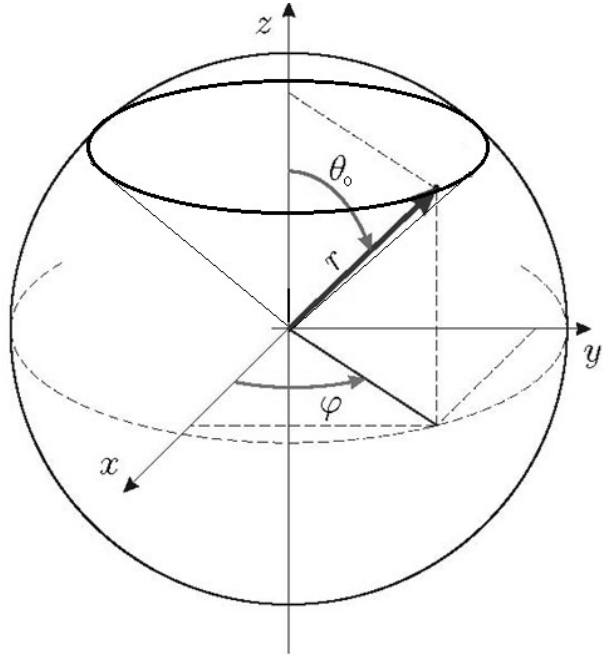

FIG. 1. A spherical coordinate system $(r,\theta,\varphi)$ and conical diffuser with polar angle $\theta_0$.

The Navier condition is a mathematical formulation of the boundary conditions of slip-with-friction, in which the tangential component of the fluid flow velocity on a solid surface is proportional to the rate of strain or, equivalently, to the viscous stress tensor component that corresponds to this rate [1-5]. So, for the surface of a conical diffuser in a spherical coordinate system [Fig. 1], the Navier condition takes the form:

$$\sigma_{\theta r}(r,\theta_0) = \frac{\eta}{\lambda} V_r(r,\theta_0), \qquad (1.6)$$

where $\lambda$ is the slip length (a constant characterizing the properties of the solid surface and the liquid), $\sigma_{\theta r}$ is the component of the viscous stress tensor corresponding to the shear strain rate [11-12]:

$$\sigma_{\theta r}(r,\theta) = \eta \left( \frac{1}{r} \frac{\partial V_r}{\partial \theta} + \frac{\partial V_\theta}{\partial r} - \frac{V_\theta}{r} \right). \qquad (1.7)$$

The second boundary condition manifests the impermeability of the conical surface:

$$V_\theta(r,\theta_0) = 0. \qquad (1.8)$$

Taking into account Eqs. (1.7) - (1.8), Eq. (1.6) can be rewritten as

$$\lambda \left( \frac{1}{r} \frac{\partial V_r(r,\theta)}{\partial \theta} + \frac{\partial V_\theta(r,\theta)}{\partial r} \right)_{\theta=\theta_0} = V_r(r,\theta_0). \qquad (1.9)$$

For more than a century, the solution of the Stokes equations (1.3) -(1.5) with no-slip conditions (1.8) -(1.9), where $\lambda = 0$, has been known [15,16]. However, for conditions of partial slip ($\lambda \neq 0$) the solution to this problem has not yet been published. One of the reasons for this situation may be as follows.

The general solution of axisymmetric Stokes equations in a spherical coordinate system, performed in the formalism of the stream function, is presented in a popular monograph [13], summarizing earlier works in which this solution was obtained [17-19]. This solution has quite complex formulation. It is difficult to use for practical problems in which the typical situation is that the boundary conditions are imposed directly on the velocity components, and not on the stream function. The disadvantage of that solution, in our opinion, is that it is based on obtaining a stream function which is not an optimal choice of the generating function, although it is very important for visualizing of two-dimensional flows.

The general solution of axisymmetric Stokes equations in a spherical coordinate system can be obtained by a mathematically equivalent, but alternative method, in the representation of a vector potential, rather than a stream function. This makes it possible to take full advantage of the well-developed apparatus of the Legendre polynomials and the attached Legendre polynomials, through which the relations for the velocity and pressure components are expressed. The solution we have obtained is shown in Table 1, and its derivation is given in the Appendix to this work.

Eqs. (A.61), (A.62), (A.76) and (A.86) of the Appendix give the general solution to the internal problem, and the expressions (A.64), (A.65), (A.79) and (A.87) are the same for the external problem. These formulas are presented in the Table 1. The solution is divided into internal and external problems for methodological purposes. However, it must be borne in mind that in the general case, the solution is the sum of the corresponding formulas for internal and external problems (left and right columns of the Table 1).

Table 1. Solutions of internal and external axisymmetric problems in a spherical coordinate system obtained in the representation of a vector potential. The radial and polar components of the incompressible fluid velocity, pressure and stream function are shown in the rows from top to bottom; $P_l$ is the Legendre polynomial, $P_l^1$ is the attached Legendre function of the first order. The derivation of the formulas is given in the Appendix.

| Internal problem | External problem |
|---|---|
| $V_r = \sum_{l=1}^{\infty} l(l+1) \left\{ \frac{a_l}{4l+6} \left(\frac{r}{R}\right)^{l+1} + c_l \left(\frac{r}{R}\right)^{l-1} \right\} P_l$ | $V_r = -\sum_{l=1}^{\infty} l(l+1) \left\{ \frac{b_l}{4l-2} \left(\frac{R}{r}\right)^{l} + d_l \left(\frac{R}{r}\right)^{l+2} \right\} P_l + d_0 \left(\frac{R}{r}\right)^{2}$ |
| $V_\theta = \sum_{l=1}^{\infty} \left\{ a_l \frac{l+3}{4l+6} \left(\frac{r}{R}\right)^{l+1} + c_l(l+1) \left(\frac{r}{R}\right)^{l-1} \right\} P_l^1$ | $V_\theta = \sum_{l=1}^{\infty} \left\{ b_l \frac{l-2}{4l-2} \left(\frac{R}{r}\right)^{l} + d_l l \left(\frac{R}{r}\right)^{l+2} \right\} P_l^1$ |
| $p = \frac{\eta}{R} \sum_{l=0}^{\infty} (l+1) a_l \left(\frac{r}{R}\right)^{l} P_l$ | $p = -\frac{\eta}{R} \sum_{l=1}^{\infty} l b_l \left(\frac{R}{r}\right)^{l+1} P_l$ |
| $\Psi(r,\theta) = -R^2 \sin\theta \sum_{l=1}^{\infty} \left\{ \frac{a_l}{4l+6} \left(\frac{r}{R}\right)^{l+3} + c_l \left(\frac{r}{R}\right)^{l+1} \right\} P_l^1(\cos\theta)$ | $\Psi(r,\theta) = R^2 \sin\theta \sum_{l=1}^{\infty} \left\{ \frac{b_l}{4l-2} \left(\frac{R}{r}\right)^{l-2} + d_l \left(\frac{R}{r}\right)^{l} \right\} P_l^1(\cos\theta) - d_0 R^2 \cos\theta$ |

Investigation of the conical diffuser problem allows to validate considered general solution in the case of no slip ($\lambda = 0$) and to obtain the new result for the boundary condition of partial slip ($\lambda \neq 0$).

## II. FLOW IN A CONICAL DIFFUSER WITH A POLAR ANGLE $0 < \theta_0 < \pi$

Let us obtain the solution of Eqs. (1.3) -(1.5) with boundary conditions given by

$$V_\theta(\infty, \theta) = V_r(\infty, \theta) = 0, \quad (2.1)$$

$$p(\infty, \theta) = 0. \quad (2.2)$$

Obviously, we have to use the formulas obtained for the external problem, which tend to zero at $r \to \infty$ (right column of Table 1).

Conditions (1.8) -(1.9) are met on the inner wall of the diffuser. Substituting $V_\theta(r, \theta_0)$ from Table 1 into condition (1.8), we obtain:

$$\sum_{l=1}^{\infty} \left\{ b_l \frac{l-2}{4l-2} \left(\frac{R}{r}\right)^{l} + d_l l \left(\frac{R}{r}\right)^{l+2} \right\} P_l^1(\cos\theta_0) = 0. \quad (2.3)$$

Similarly, substituting $V_r(r, \theta)$, one can find the left side of Eq. (1.9)

$$\left( \frac{1}{r} \frac{\partial V_r(r,\theta)}{\partial \theta} + \frac{\partial V_\theta(r,\theta)}{\partial r} \right)_{\theta=\theta_0} = -\frac{1}{R} \sum_{l=1}^{\infty} l \left\{ \frac{b_l}{2} \left(\frac{R}{r}\right)^{l+1} + d_l(2l+3) \left(\frac{R}{r}\right)^{l+3} \right\} P_l^1(\cos\theta_0), \quad (2.4)$$

where the transformations take into account the following expression [20]

$$\frac{\partial P_l}{\partial \theta} = -\sin\theta \frac{\partial P_l}{\partial \cos\theta} = P_l^1. \tag{2.5}$$

Substituting Eq. (2.4) and $V_r(r,\theta)$ from the Table 1 in Eq. (1.9), one can obtain

$$\frac{\lambda}{R}\sum_{l=1}^{\infty} l \left\{ \frac{b_l}{2}\left(\frac{R}{r}\right)^{l+1} + d_l(2l+3)\left(\frac{R}{r}\right)^{l+3} \right\} P_l^1(\cos\theta_0) =$$
$$= \sum_{l=1}^{\infty} l(l+1)\left\{ \frac{b_l}{4l-2}\left(\frac{R}{r}\right)^{l} + d_l\left(\frac{R}{r}\right)^{l+2} \right\} P_l(\cos\theta_0) - d_0\left(\frac{R}{r}\right)^2. \tag{2.6}$$

There are two terms under summation in Eq. (2.3). The first of them can be represented as

$$\sum_{l=1}^{\infty} b_l \frac{l-2}{4l-2}\left(\frac{R}{r}\right)^l P_l^1(\cos\theta_0) = -b_1 \frac{R}{2r} P_1^1(\cos\theta_0) + \sum_{l=3}^{\infty} b_l \frac{l-2}{4l-2}\left(\frac{R}{r}\right)^l P_l^1(\cos\theta_0). \tag{2.7}$$

Shifting the count of summation in $l$ by two units downwards in the right part of the equation (2.7) and making a redefinition, we get:

$$\sum_{l=1}^{\infty} b_l \frac{l-2}{4l-2}\left(\frac{R}{r}\right)^l P_l^1(\cos\theta_0) = -b_1 \frac{R}{2r} P_1^1(\cos\theta_0) + \sum_{l=1}^{\infty} b_{l+2} \frac{l}{4l+6}\left(\frac{R}{r}\right)^{l+2} P_{l+2}^1(\cos\theta_0). \tag{2.8}$$

Substituting Eq. (2.8) in Eq. (2.3), we have

$$-b_1\frac{R}{2r}P_1^1(\cos\theta_0) + \sum_{l=1}^{\infty} l\left\{ \frac{b_{l+2}}{4l+6}P_{l+2}^1(\cos\theta_0) + d_l P_l^1(\cos\theta_0) \right\}\left(\frac{R}{r}\right)^{l+2} = 0. \tag{2.9}$$

Therefore

$$b_1 P_1^1(\cos\theta_0) = 0, \quad \frac{b_{l+2}}{4l+6}P_{l+2}^1(\cos\theta_0) + d_l P_l^1(\cos\theta_0) = 0, \ l=1,2,3, \ldots \tag{2.10}$$

Similarly, let us convert the sum on the right side of the condition (2.6):

$$\sum_{l=1}^{\infty} l(l+1)\left\{ \frac{b_l}{4l-2}\left(\frac{R}{r}\right)^l + d_l\left(\frac{R}{r}\right)^{l+2} \right\} P_l(\cos\theta_0) = \left\{ \frac{b_1 R}{r} + 2d_1\left(\frac{R}{r}\right)^3 \right\} P_1(\cos\theta_0) +$$
$$+ \sum_{l=1}^{\infty} (l+1)(l+2)\left\{ \frac{b_{l+1}}{4l+2}\left(\frac{R}{r}\right)^{l+1} + d_{l+1}\left(\frac{R}{r}\right)^{l+3} \right\} P_{l+1}(\cos\theta_0). \tag{2.11}$$

Substituting Eq. (2.12) in Eq. (2.6), one can obtain

$$\frac{\lambda}{R}\sum_{l=1}^{\infty} l\left\{ \frac{b_l}{2}\left(\frac{R}{r}\right)^{l+1} + d_l(2l+3)\left(\frac{R}{r}\right)^{l+3} \right\} P_l^1(\cos\theta_0) - \sum_{l=1}^{\infty}(l+1)(l+2)\left\{ \frac{b_{l+1}}{4l+2}\left(\frac{R}{r}\right)^{l+1} + d_{l+1}\left(\frac{R}{r}\right)^{l+3} \right\} P_{l+1}(\cos\theta_0) =$$
$$= \left\{ \frac{b_1 R}{r} + 2d_1\left(\frac{R}{r}\right)^3 \right\} P_1(\cos\theta_0) - d_0\left(\frac{R}{r}\right)^2. \tag{2.12}$$

Rearranging and combining the terms by equal powers of $\dfrac{R}{r}$, we have

$$\sum_{l=1}^{\infty}\left\{b_l l \frac{\lambda}{2R} P_l^1(\cos\theta_0) - (l+1)(l+2)\frac{b_{l+1}}{4l+2} P_{l+1}(\cos\theta_0)\right\}\left(\frac{R}{r}\right)^{l+1} +$$

$$+\sum_{l=1}^{\infty}\left\{d_l \frac{(2l+3)l\lambda}{R} P_l^1(\cos\theta_0) - (l+1)(l+2)d_{l+1}P_{l+1}(\cos\theta_0)\right\}\left(\frac{R}{r}\right)^{l+3} = \qquad (2.13)$$

$$=\left\{\frac{b_1 R}{r} + 2d_1\left(\frac{R}{r}\right)^3\right\}P_1(\cos\theta_0) - d_0\left(\frac{R}{r}\right)^2.$$

Let us select the first two terms in the first sum in Eq. (2.13), bring the rest to the system of counting powers of $\dfrac{R}{r}$ that corresponds to the second sum, and shift the counting along $l$:

$$\sum_{l=1}^{2}\left\{b_l l \frac{\lambda}{2R} P_l^1(\cos\theta_0) - (l+1)(l+2)\frac{b_{l+1}}{4l+2} P_{l+1}(\cos\theta_0)\right\}\left(\frac{R}{r}\right)^{l+1} +$$

$$+\sum_{l=1}^{\infty}\left\{b_{l+2}(l+2)\frac{\lambda}{2R} P_{l+2}^1(\cos\theta_0) - (l+3)(l+4)\frac{b_{l+3}}{4l+10} P_{l+3}(\cos\theta_0)\right\}\left(\frac{R}{r}\right)^{l+3} +$$

$$+\sum_{l=1}^{\infty}\left\{d_l \frac{(2l+3)l\lambda}{R} P_l^1(\cos\theta_0) - (l+1)(l+2)d_{l+1}P_{l+1}(\cos\theta_0)\right\}\left(\frac{R}{r}\right)^{l+3} = \qquad (2.14)$$

$$=\left\{\frac{b_1 R}{r} + 2d_1\left(\frac{R}{r}\right)^3\right\}P_1(\cos\theta_0) - d_0\left(\frac{R}{r}\right)^2.$$

Coefficients of equal powers of $\dfrac{R}{r}$ can be combined:

$$\sum_{l=1}^{\infty}\left\{b_{l+2}(l+2)\frac{\lambda}{2R}P_{l+2}^1(\cos\theta_0) + d_l\frac{(2l+3)l\lambda}{R}P_l^1(\cos\theta_0) - (l+3)(l+4)\frac{b_{l+3}}{4l+10}P_{l+3}(\cos\theta_0) - (l+1)(l+2)d_{l+1}P_{l+1}(\cos\theta_0)\right\}\left(\frac{R}{r}\right)^{l+3} =$$

$$=\frac{b_1 R}{r}P_1(\cos\theta_0)+\left\{b_2 P_2(\cos\theta_0) - d_0 - b_1\frac{\lambda}{2R}P_1^1(\cos\theta_0)\right\}\left(\frac{R}{r}\right)^2 + \left\{2d_1 P_1(\cos\theta_0) - b_2\frac{\lambda}{R}P_2^1(\cos\theta_0) + \frac{6b_3}{5}P_3(\cos\theta_0)\right\}\left(\frac{R}{r}\right)^3.$$

(2.15)

To satisfy Eq. (2.15), it is necessary to require the following conditions

$$b_1 P_1(\cos\theta_0) = 0, \qquad (2.16)$$

$$b_2 P_2(\cos\theta_0) - d_0 - b_1 \frac{\lambda}{2R} P_1^1(\cos\theta_0) = 0, \qquad (2.17)$$

$$2d_1 P_1(\cos\theta_0) - b_2 \frac{\lambda}{R} P_2^1(\cos\theta_0) + \frac{6b_3}{5} P_3(\cos\theta_0) = 0, \qquad (2.18)$$

$$b_{l+2}(l+2)\frac{\lambda}{2R} P_{l+2}^1(\cos\theta_0) + d_l \frac{(2l+3)l\lambda}{R} P_l^1(\cos\theta_0) -$$

$$-(l+3)(l+4)\frac{b_{l+3}}{4l+10} P_{l+3}(\cos\theta_0) - (l+1)(l+2)d_{l+1}P_{l+1}(\cos\theta_0) = 0, \quad l=1,2,3,... \qquad (2.19)$$

Let us assume that
$$P_l \neq 0 \text{ and } P_l^1 \neq 0. \tag{2.20}$$

Conditions (2.10) can be conveniently rewritten as:

$$b_1 = 0, \quad d_l = -\frac{P_{l+2}^1(\cos\theta_0)}{P_l^1(\cos\theta_0)} \frac{b_{l+2}}{4l+6}, \quad l=1,2,3,\ldots \tag{2.21}$$

Taking this into account, condition (2.17) can be rewritten as:

$$d_0 = b_2 P_2(\cos\theta_0), \tag{2.22}$$

and conditions (2.18) -(2.19) can be combined by the formula

$$b_{l+2} = b_{l+1}(4l+6)\frac{\lambda}{R}P_{l+1}^1(\cos\theta_0)\left\{(l+2)(l+3)P_{l+2}(\cos\theta_0)-l(l+1)\frac{P_{l+2}^1(\cos\theta_0)}{P_l^1(\cos\theta_0)}P_l(\cos\theta_0)\right\}^{-1}, \quad l=1,2,3\ldots \tag{2.23}$$

Eq. (2.23) is a recurrence relation that allows us to sequentially calculate $b_3$, $b_4$, etc., if the coefficient $b_2$ is known. In particular, we have:

$$b_3 = 5b_2 \frac{\lambda}{R} P_2^1(\cos\theta_0)\left[6P_3(\cos\theta_0)-P_1(\cos\theta_0)\frac{P_3^1(\cos\theta_0)}{P_1^1(\cos\theta_0)}\right]^{-1}, \tag{2.24}$$

$$b_4 = 7b_3 \frac{\lambda}{R} P_3^1(\cos\theta_0)\left\{10P_4(\cos\theta_0)-3\frac{P_4^1(\cos\theta_0)}{P_2^1(\cos\theta_0)}P_2(\cos\theta_0)\right\}^{-1}. \tag{2.25}$$

The corresponding coefficients $d_l$ are determined by Eq. (2.21). In particular,

$$d_1 = -\frac{P_3^1(\cos\theta_0)}{P_1^1(\cos\theta_0)}P_2^1(\cos\theta_0)\frac{\lambda b_2}{2R}\left(6P_3(\cos\theta_0)-P_1(\cos\theta_0)\frac{P_3^1(\cos\theta_0)}{P_1^1(\cos\theta_0)}\right)^{-1}. \tag{2.26}$$

The flow rate of liquid passing through the diffuser is determined by the formula:

$$Q = 2\pi r^2 \int_0^{\theta_0} V_r(r,\theta)\sin\theta d\theta. \tag{2.27}$$

The flow rate must be constant in any section of the cone: $Q = \text{const}$. Then

$$\int_0^{\theta_0} V_r(r,\theta)\sin\theta d\theta \sim \frac{1}{r^2}. \tag{2.28}$$

Let us verify that the constancy of the flow rate across the cone cross-section actually takes place.

Substituting $V_r$ from Table 1 in Eq. (2.27), one can obtain

$$Q = 2\pi r^2 \sum_{l=1}^{\infty} l(l+1)\left\{\frac{b_l}{4l-2}\left(\frac{R}{r}\right)^l + d_l\left(\frac{R}{r}\right)^{l+2}\right\} \int_1^{\cos\theta_0} P_l(t)dt + 2\pi R^2 d_0 \int_0^{\theta_0} \sin\theta d\theta, \tag{2.29}$$

where $\int_1^{\cos\theta_0} P_l(t)dt$, in general, are some non-zero constants.

Let us break the sum on the right side of Eq. (2.29) into terms as follows

$$\sum_{l=1}^{\infty} l(l+1) \left\{ \frac{b_l}{4l-2} \left(\frac{R}{r}\right)^l + d_l \left(\frac{R}{r}\right)^{l+2} \right\} \int_1^{\cos\theta_0} P_l(t)dt = \frac{b_1 R}{r} \int_1^{\cos\theta_0} P_1(t)dt + b_2 \left(\frac{R}{r}\right)^2 \int_1^{\cos\theta_0} P_2(t)dt +$$
$$+ \sum_{l=1}^{\infty} (l+2)(l+3) \frac{b_{l+2}}{4l+6} \left(\frac{R}{r}\right)^{l+2} \int_1^{\cos\theta_0} P_{l+2}(t)dt + \sum_{l=1}^{\infty} l(l+1) d_l \left(\frac{R}{r}\right)^{l+2} \int_1^{\cos\theta_0} P_l(t)dt \quad (2.30)$$

In the penultimate sum on the right side of Eq. (2.30) the starting point for $l$ is shifted from 3 to 1 to synchronize it with the last sum. Combining the sums, we have

$$\sum_{l=1}^{\infty} l(l+1) \left\{ \frac{b_l}{4l-2} \left(\frac{R}{r}\right)^l + d_l \left(\frac{R}{r}\right)^{l+2} \right\} \int_1^{\cos\theta_0} P_l(t)dt = \frac{b_1 R}{r} \int_1^{\cos\theta_0} P_1(t)dt + b_2 \left(\frac{R}{r}\right)^2 \int_1^{\cos\theta_0} P_1(t)dt +$$
$$+ \sum_{l=1}^{\infty} \left\{ (l+2)(l+3) \frac{b_{l+2}}{4l+6} \int_1^{\cos\theta_0} P_{l+2}(t)dt + l(l+1) d_l \int_1^{\cos\theta_0} P_l(t)dt \right\} \left(\frac{R}{r}\right)^{l+2} \quad (2.31)$$

Obviously, to ensure Eq. (2.28), the following relations must be satisfied:

$$b_1 \int_1^{\cos\theta_0} P_1(t)dt = 0, \quad (2.32)$$

$$(l+2)(l+3) \frac{b_{l+2}}{4l+6} \int_1^{\cos\theta_0} P_{l+2}(t)dt + l(l+1) d_l \int_1^{\cos\theta_0} P_l(t)dt = 0. \quad (2.33)$$

To calculate $\int_1^{\cos\theta_0} P_l(t)dt$, one can use the differential equation for Legendre polynomials [20]:

$$\frac{d}{dt}\left[(1-t^2)\frac{dP_l(t)}{dt}\right] = -l(l+1)P_l(t). \quad (2.34)$$

Therefore

$$\int_1^{\cos\theta_0} P_l(t)dt = -\frac{(1-t^2)}{l(l+1)} \frac{dP_l(t)}{dt} \bigg|_1^{\cos\theta_0}. \quad (2.35)$$

Let us use the definition of the first-order associated Legendre polynomials [20]:

$$P_l^1(t) = -\sqrt{1-t^2}\,\frac{dP_l(t)}{dt}. \quad (2.36)$$

Substituting Eq. (2.36) in Eq. (2.35), one can obtain

$$\int_1^{\cos\theta_0} P_l(t)dt = \frac{\sqrt{1-t^2}}{l(l+1)} P_l^1(t) \bigg|_1^{\cos\theta_0} = \frac{\sin\theta_0}{l(l+1)} P_l^1(\cos\theta_0). \quad (2.37)$$

Substituting Eq. (2.37) into conditions (2.32) and (2.33), we obtain (for $\sin\theta_0 \neq 0$) exactly the conditions (2.10), which were already taken into account in solving the problem and mean the impermeability of the cone wall. Hence Eq. (2.30) takes the form

$$\sum_{l=1}^{\infty} l(l+1) \left\{ \frac{b_l}{4l-2} \left(\frac{R}{r}\right)^l + d_l \left(\frac{R}{r}\right)^{l+2} \right\} \int_1^{\cos\theta_0} P_l(t)dt = b_2 \left(\frac{R}{r}\right)^2 \int_1^{\cos\theta_0} P_2(t)dt. \quad (2.38)$$

Substituting Eq. (2.38) into Eq. (2.29), one can get

$$Q = 2\pi R^2 b_2 \int_1^{\cos\theta_0} P_2(t)dt + 2\pi R^2 d_0 \int_0^{\theta_0} \sin\theta d\theta. \qquad (2.39)$$

Carrying out calculations and substituting Eq. (2.22), we obtain

$$Q = \pi R^2 b_2 \left[3\cos^2\theta_0 - 2\cos^3\theta_0 - 1\right] = -\pi R^2 b_2 (1-\cos\theta_0)^2 (1+2\cos\theta_0). \qquad (2.40)$$

At $\cos\theta_0 = -\frac{1}{2}$ ($\theta_0 = \frac{2\pi}{3}$) the flow rate given by Eq.(2.40) becomes zero, so that the flow satisfying the boundary conditions is also zero, and it is impossible to determine the coefficient $b_2$. For all other cases, one can write

$$b_2 = -\frac{Q}{\pi R^2 (1-\cos\theta_0)^2 (1+2\cos\theta_0)}. \qquad (2.41)$$

Since all other coefficients are sequentially calculated through $b_2$, the solution can be expressed in terms of the flow rate.

If the parameter $\frac{\lambda}{R}$ is a small value, then the coefficients $b_2$, $b_3$, $b_4$, etc., as follows from Eq. (2.23), form a series, each term of which has a higher order of smallness than the previous one. We find the projections of the velocity, pressure and stream function with an accuracy of up to the first power of the parameter $\frac{\lambda}{R}$. For this, as indicated by Eqs. (2.14) and (2.16), it is sufficient to limit ourselves to taking into account the terms with the coefficients $b_2$, $b_3$ and $d_1$. According to the formulas presented in Table 1, in the indicated approximation, the components of the velocity, pressure and stream function have the form

$$V_r \approx -\left(2d_1 P_1(\cos\theta) + \frac{6b_3}{5} P_3(\cos\theta)\right)\left(\frac{R}{r}\right)^3 + b_2 (P_2(\cos\theta_0) - P_2(\cos\theta))\left(\frac{R}{r}\right)^2, \qquad (2.42)$$

$$V_\theta \approx \left(d_1 P_1^1 + \frac{b_3}{10} P_3^1\right)\left(\frac{R}{r}\right)^3, \qquad (2.43)$$

$$p = -\frac{2\eta}{R} b_2 \left(\frac{R}{r}\right)^3 P_2 - \frac{3\eta}{R} b_3 \left(\frac{R}{r}\right)^4 P_3, \qquad (2.44)$$

$$\Psi(r,\theta) \approx b_2 \frac{R^2}{2} \cos\theta \left\{\cos^2\theta - 3\cos^2\theta_0\right\} + \frac{R^3}{r} \sin\theta \left\{d_1 P_1^1(\cos\theta) + \frac{b_3}{10} P_3^1(\cos\theta)\right\}. \qquad (2.45)$$

Taking into account $P_1(t) = t$, $P_2(t) = \frac{1}{2}(3t^2 - 1)$, $P_3(t) = \frac{1}{2}(5t^3 - 3t)$, $P_1^1(t) = -\sqrt{1-t^2}$,

$P_2^1(t) = -3t\sqrt{1-t^2}$, and $P_3^1(t) = -\frac{3}{2}(5t^2 - 1)\sqrt{1-t^2}$ [20], Eqs. (2.24) and (2.26) give:

$$b_3 = \frac{2\lambda}{R}\frac{b_2}{\sin\theta_0}, \tag{2.46}$$

$$d_1 = -\frac{3}{20}b_3(5\cos^2\theta_0 - 1). \tag{2.47}$$

Substituting Eqs. (2.46) and (2.47) in Eqs. (2.42)- (2.45), one can obtain the desired solution of the problem

$$V_r \approx \frac{3b_2}{2}(\cos^2\theta_0 - \cos^2\theta)\left(\frac{R}{r}\right)^2 + \frac{3b_2\lambda}{R}\frac{\cos\theta}{\sin\theta_0}(\cos^2\theta_0 - 2\cos^2\theta + 1)\left(\frac{R}{r}\right)^3, \tag{2.48}$$

$$V_\theta \approx \frac{3b_2\lambda}{2R}\frac{\sin\theta}{\sin\theta_0}(\cos^2\theta_0 - \cos^2\theta)\left(\frac{R}{r}\right)^3, \tag{2.49}$$

$$p \approx -\frac{\eta}{R}b_2\left(\frac{R}{r}\right)^3(3\cos^2\theta - 1) - \frac{3\lambda\eta}{R^2}\frac{b_2}{\sin\theta_0}\left(\frac{R}{r}\right)^4(5\cos^2\theta - 3)\cos\theta, \tag{2.50}$$

$$\Psi(r,\theta) \approx \frac{b_2 R^2}{2}\cos\theta(\cos^2\theta - 3\cos^2\theta_0) + \frac{3\lambda b_2 R^2}{2r}\frac{\sin^2\theta}{\sin\theta_0}(\cos^2\theta_0 - \cos^2\theta). \tag{2.51}$$

Let us express Eqs. (2.48)-(2.51) through the liquid flow rate $Q$ by substituting Eq. (2.41):

$$V_r \approx 3Q\frac{\cos^2\theta - \cos^2\theta_0 - \frac{2\lambda}{r}\frac{\cos\theta}{\sin\theta_0}(\cos^2\theta_0 - 2\cos^2\theta + 1)}{2\pi r^2(1-\cos\theta_0)^2(1+2\cos\theta_0)}, \tag{2.52}$$

$$V_\theta \approx \frac{3\lambda Q \sin\theta(\cos^2\theta - \cos^2\theta_0)}{2\pi r^3 \sin\theta_0(1-\cos\theta_0)^2(1+2\cos\theta_0)}, \tag{2.53}$$

$$p \approx \eta Q\frac{3\cos^2\theta - 1 + \frac{3\lambda}{r}\frac{\cos\theta}{\sin\theta_0}(5\cos^2\theta - 3)}{\pi r^3(1-\cos\theta_0)^2(1+2\cos\theta_0)}, \tag{2.54}$$

$$\Psi(r,\theta) \approx Q\frac{\cos\theta(3\cos^2\theta_0 - \cos^2\theta) + \frac{3\lambda}{r}\frac{\sin^2\theta}{\sin\theta_0}(\cos^2\theta - \cos^2\theta_0)}{2\pi(1-\cos\theta_0)^2(1+2\cos\theta_0)}. \tag{2.55}$$

### III. DISCUSSION

The following condition must be satisfied in Eqs. (2.52) -(2.55)

$$\frac{\lambda}{r} \ll 1. \tag{3.1}$$

Indeed, the condition of applicability of Eqs. (2.48) -(2.51) is $\lambda \ll R$ or, due to

$$\frac{\lambda}{R}\left(\frac{R}{r}\right)^n = \frac{\lambda}{r}\left(\frac{R}{r}\right)^{n-1}, \quad n = 1, 2, 3, \ldots, \tag{3.2}$$

this condition is practically equivalent to $r \gg \lambda$, i.e. the solution is applicable in the region of large distances from the cone apex and/or small slip lengths. Thus, parameter $R$ plays the role of a characteristic unit of measurement of distance in space.

Note that to ensure the validity of the Stokes equations, it is also necessary to consider the region of the cone that is sufficiently far from its apex. Indeed, the Stokes equations are valid for small Reynolds numbers:

$$\text{Re} = \frac{\rho L V}{\eta} \ll 1, \tag{3.3}$$

where $\rho$ is the density of the liquid, $V$ is the characteristic velocity, $L$ is the characteristic size, which for a cone can be considered its cutting radius at a given $r$. In fact, keeping in mind the approximation (3.3), the characteristic size is the radial coordinate $r$, measured from the cone apex, which in the most interesting cases has the same order of magnitude as the corresponding radius of the cone funnel for a given $r$. The ratio of the flow rate to the square of the characteristic size $V \propto \frac{Q}{r^2}$ can serve as an estimate for the velocity. Therefore, instead of Eq. (3.3), we can write

$$r \gg \frac{\rho Q}{\eta}. \tag{3.4}$$

In other words, both the Stokes equations and their linear approximation in the expansion in $\lambda$ work better the further from the cone apex the area of study is. But, generally speaking, the ratio of the parameters $\frac{\rho Q}{\eta}$ and $\lambda$ can have any value.

For $\lambda = 0$, taking into account Eqs. (2.52)-(2.55), one can obtain the well-known exact solution of the Stokes equations for the no slip condition [13,15,16]:

$$V_r = \frac{3Q(\cos^2\theta - \cos^2\theta_0)}{2\pi r^2 (1-\cos\theta_0)^2(1+2\cos\theta_0)}, \tag{3.5}$$

$$V_\theta = 0, \tag{3.6}$$

$$p = \frac{\eta Q(3\cos^2\theta - 1)}{\pi r^3 (1-\cos\theta_0)^2(1+2\cos\theta_0)}, \tag{3.7}$$

$$\Psi(r,\theta) = \frac{Q\cos\theta(3\cos^2\theta_0 - \cos^2\theta)}{2\pi(1-\cos\theta_0)^2(1+2\cos\theta_0)}. \tag{3.8}$$

The streamlines constructed on the basis of Eqs. (2.48), (2.49) and (2.51) are shown in Fig. 2. The following parameter values were adopted in the calculations: $b_2 = -\frac{1}{3}$, $R = 1$. Fig. 2 shows that a non-zero slip length ($\lambda > 0$) leads to the occurrence of a vortex of the liquid flow, which increases with increasing parameter $\lambda$. The qualitative difference between the flow at $\lambda > 0$ and the flow at $\lambda = 0$ is that in the first case the polar component of velocity, $V_\theta$, is non-zero, while in the second case Eq. (3.6) is satisfied, so that the liquid flows strictly radially from the apex.

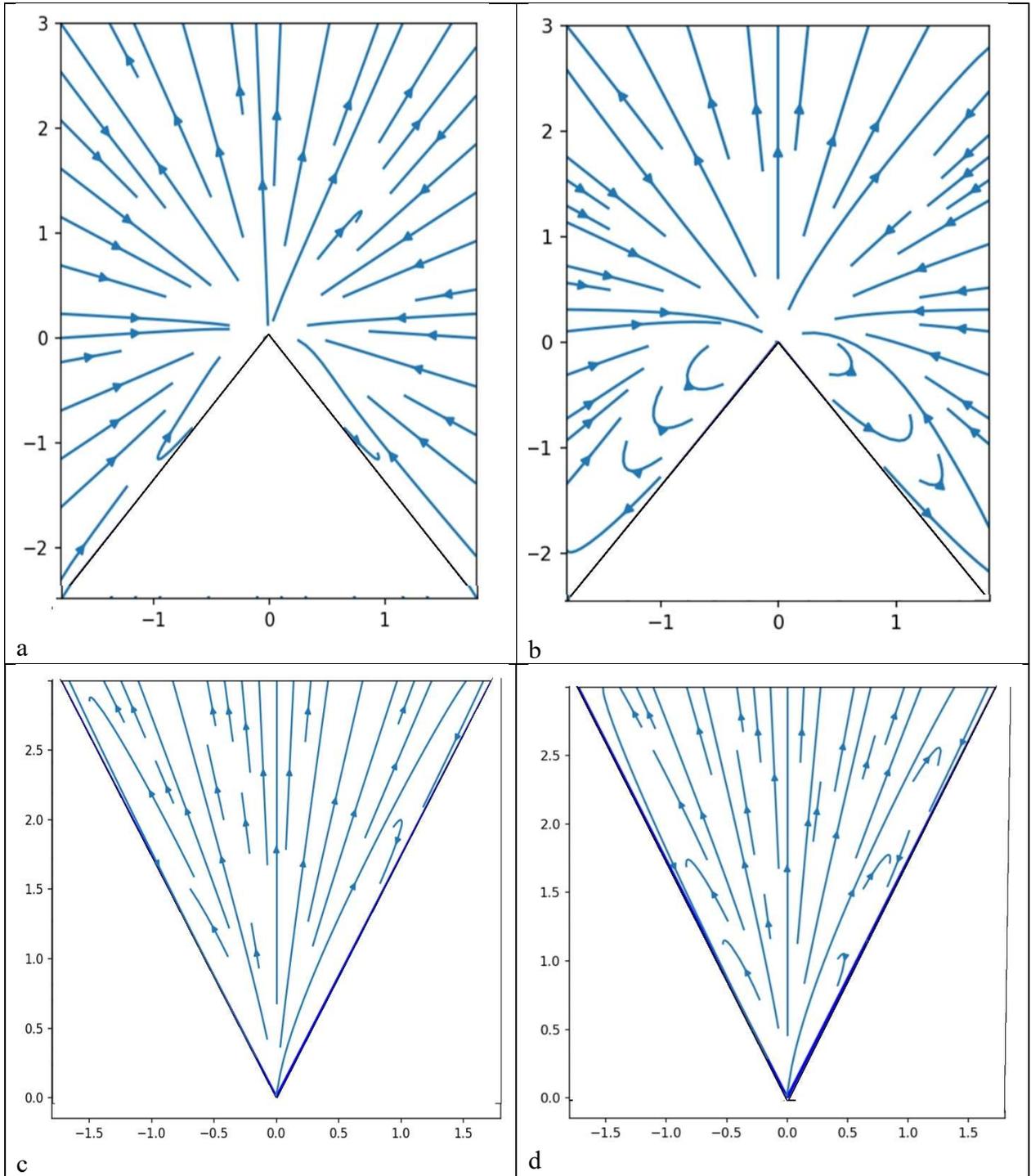

FIG. 2. Streamlines in a cone for different values of the polar angle and slip length: a) $\frac{\lambda}{R} = 0.075$, $\theta_0 = \frac{4\pi}{5}$; b) $\frac{\lambda}{R} = 0.2$, $\theta_0 = \frac{4\pi}{5}$; c) $\frac{\lambda}{R} = 0.2$, $\theta_0 = \frac{\pi}{6}$; d) $\frac{\lambda}{R} = 0.3$, $\theta_0 = \frac{\pi}{6}$. Axes scale in dimensionless units $\frac{r}{R}$.

To obtain a solution suitable for $r \sim \lambda$, it is necessary to take into account the following terms in the expansion of the velocity and pressure in a series in $\frac{\lambda}{R}$. To do this, it is necessary to use the recurrence relations (2.21) - (2.23) to consequentially determine the coefficients $b_l$ and $d_l$, corresponding to increasingly higher values of *l*.

**IV. CONCLUSION**

The problem of evaporation of a sessile droplet was solved earlier using a mathematical analogy with electrostatics [21-24]. Similarly, the Stokes equations can be solved with partial use of an analogy with direct current magnetism, where the vector potential formalism is used. Such a solution method is given in the Appendix. In this case, the stream function can be calculated as a secondary auxiliary quantity. On the contrary, in the existing generally accepted approach, going back to Stokes, the main sought-for quantity is the stream function [12,13].

An alternative form of the general solution of the linearized stationary axisymmetric Navier–Stokes equations for an incompressible fluid in spherical coordinates is obtained (Table 1). The previously published solution of this problem [13, 17-19] is given in terms of the stream function, which leads to formulas that are quite complex for practical application. From a mathematical point of view, it is more "natural" to use the representation of velocity through a vector potential, since this allows one to directly apply the well-developed apparatus of polynomials and associated Legendre functions. At the same time, both approaches are mathematically equivalent, which is proven by both general considerations and specific calculations.

The new form of the solution is applied to the problem of fluid flow through a conical diffuser under boundary conditions of partial slip for an arbitrary slip length. In particular, for $\lambda = 0$ we obtain the limiting case, the well-known solution by Harrison to the problem of a diffuser with no slip boundary conditions [12,13,15,16].

The considered example shows that the solution given in Table 1 and obtained in the Appendix to this work allows to successfully solve quite complex boundary problems. It is easy to verify that the proposed general solution, when substituted into the corresponding boundary conditions, allows to obtain correct expressions for such test cases as the Stokes drag force, the Hadamard–Rybczynski equation (describing the motion of a spherical drop of liquid in another external liquid), etc. Successful verification of these equations allows Table 1 to be recommended for further use.

# APPENDIX: SOLUTION OF THE AXISYMMETRIC STOKES PROBLEM IN THE REPRESENTATION OF THE VECTOR POTENTIAL

The vector Laplace operator is introduced as follows [14]:
$$\vec{\nabla}^2 \mathbf{A} = \nabla(\nabla \cdot \mathbf{A}) - \nabla \times [\nabla \times \mathbf{A}], \qquad (A.1)$$
while the scalar Laplace operator has the form:
$$\nabla^2 f = \nabla(\nabla f). \qquad (A.2)$$

It is known that the operators $\vec{\nabla}^2$ and $\nabla^2$ are identically equal in rectangular Cartesian coordinates, but differ in curvilinear coordinates.

In the case of small Reynolds numbers, the steady flow of a viscous incompressible fluid is described by the linearized Navier–Stokes equations (or Stokes equations) [12]:
$$\eta \vec{\nabla}^2 \mathbf{V} = \nabla p, \qquad (A.3)$$
$$(\nabla \cdot \mathbf{V}) = 0. \qquad (A.4)$$

In rectangular Cartesian coordinates, consecutive operators $\nabla$ can be interchanged, but in curvilinear coordinates this is generally not the case. Next, we obtain the basic expressions using the Cartesian system.

If we apply a vector operator $\nabla \times$ (curl) to equation (A.3), written in Cartesian coordinates, and change the order of differentiation, we have [11-13]
$$\vec{\nabla}^2 [\nabla \times \mathbf{V}] = 0. \qquad (A.5)$$
Applying the divergence operator $\nabla \cdot$ to the same equation, taking into account (A.4), we obtain:
$$\nabla^2 p = 0. \qquad (A.6)$$

Let us introduce the velocity circulation vector (vorticity)
$$\mathbf{\Phi} = [\nabla \times \mathbf{V}]. \qquad (A.7)$$
Then (A.5) can be rewritten as

$$\vec{\nabla}^2 \mathbf{\Phi} = 0. \qquad (A.8)$$

Any vector field, for example, the velocity vector field, can be represented as the sum of the gradient of some scalar function $f$ and the vortex part described by the curl of some vector $\mathbf{A}$:
$$\mathbf{V} = \nabla f + [\nabla \times \mathbf{A}]. \qquad (A.9)$$
In this case, from the continuity equation (A.4) it follows:
$$\nabla^2 f = 0. \qquad (A.10)$$
Thus, the divergence of both velocity terms in Eq. (A.9) must be zero not only in sum, but also separately. Representation (A.9) corresponds to the division of velocity into irrotational (longitudinal) and vortex (transverse) components:
$$\mathbf{V} = \mathbf{V}_\| + \mathbf{V}_\perp, \qquad (A.11)$$
where
$$\mathbf{V}_\| = \nabla f, \qquad (A.12)$$

$$\mathbf{V}_\perp = [\nabla \times \mathbf{A}]. \qquad (A.13)$$
The following conditions take place:
$$(\nabla \cdot \mathbf{V}_\|) = (\nabla \cdot \mathbf{V}_\perp) = 0. \qquad (A.14)$$
Therefore, relation (A.7) can be refined:
$$\mathbf{\Phi} = [\nabla \times \mathbf{V}_\perp], \qquad (A.15)$$
$$[\nabla \times \mathbf{V}_\|] = 0. \qquad (A.16)$$

Substituting Eq. (A.11) into Eq. (A.3), taking into account Eq. (A.10), one can obtain an important clarification of Eqs. (A.3)

$$\eta \overrightarrow{\nabla^2 \mathbf{V}_\perp} = \nabla p, \qquad (A.17)$$

$$\overrightarrow{\nabla^2 \mathbf{V}_\parallel} = 0, \qquad (A.18)$$

so that the pressure gradient is determined only by the vortex part of the velocity field.

### A.1. Calculation of vorticity

Consider the axisymmetric problem for Eqs. (A.3) -(A.4) in spherical coordinates [Fig. 1]:

$$\nabla^2 V_r - \frac{2V_r}{r^2} - \frac{2}{r^2 \sin\theta} \frac{\partial}{\partial \theta}(V_\theta \sin\theta) = \eta^{-1} \frac{\partial}{\partial r} p, \qquad (A.19)$$

$$\nabla^2 V_\theta - \frac{V_\theta}{r^2 \sin^2\theta} + \frac{2}{r^2} \frac{\partial}{\partial \theta} V_r = \eta^{-1} \frac{1}{r} \frac{\partial}{\partial \theta} p, \qquad (A.20)$$

$$\frac{1}{r} \frac{\partial}{\partial r}(r^2 V_r) + \frac{1}{\sin\theta} \frac{\partial}{\partial \theta}(V_\theta \sin\theta) = 0, \qquad (A.21)$$

where the non-zero components of velocity $V_r(r,\theta)$, $V_\theta(r,\theta)$ and pressure $p(r,\theta)$ do not depend on the azimuthal angle φ.

In the axisymmetric case, there is only one non-zero component of the vorticity vector, namely the φ-component, which we will denote by $\Phi$. In this case, equation (A.8) takes the form

$$\nabla^2_\varphi \Phi = 0, \qquad (A.22)$$

where $\nabla^2_\varphi$ is the φ-component of vector Laplacian $\overrightarrow{\nabla^2}$ that is written in the spherical coordinate system as follows

$$\nabla^2_\varphi = \frac{1}{r^2} \frac{\partial}{\partial r}\left(r^2 \frac{\partial}{\partial r}\right) + \frac{1}{r^2 \sin\theta} \frac{\partial}{\partial \theta}\left(\sin\theta \frac{\partial}{\partial \theta}\right) - \frac{1}{r^2 \sin^2\theta}, \qquad (A.23)$$

For the internal problem ($0 \leq r \leq R$) the solution of Eq. (A.22) has the form [20]:

$$\Phi(r,\theta) = \sum_{l=1}^{\infty} A_l \left(\frac{r}{R}\right)^l P_l^1(\cos\theta). \qquad (A.24)$$

Similarly, for the external problem we have

$$\Phi(r,\theta) = \sum_{l=1}^{\infty} B_l \left(\frac{R}{r}\right)^{l+1} P_l^1(\cos\theta), \qquad (A.25)$$

where $P_l^1$ is the associated Legendre polynomial of the first order, $A_l$ and $B_l$ are the coefficients.

The vector potential of the transverse component of the velocity has only the φ-component, which is denoted as $A$. Then, using the rules for calculating the curl in spherical coordinates, one can calculate both non-zero components of the velocity:

$$V_{\perp r} = \frac{1}{r \sin\theta} \frac{\partial}{\partial \theta}[A(r,\theta) \sin\theta], \qquad (A.26)$$

$$V_{\perp \theta} = -\frac{1}{r} \frac{\partial}{\partial r}[r A(r,\theta)]. \qquad (A.27)$$

## A.2. Calculation of vector potential

The relationship between vorticity and vector potential follows from Eqs. (A.7) and (A.13):
$$\Phi = -\nabla_\varphi^2 A. \qquad (A.28)$$

Then Eq. (A.22) can be rewritten as:
$$\nabla_\varphi^2 \nabla_\varphi^2 A = 0. \qquad (A.29)$$

Taking into account Eq. (A.28), for the internal problem we have
$$\nabla_\varphi^2 A = -\sum_{l=1}^{\infty} A_l \left(\frac{r}{R}\right)^l P_l^1(\cos\theta) \qquad (A.30)$$

and a similar equation for the external problem
$$\nabla_\varphi^2 A = -\sum_{l=1}^{\infty} A_l \left(\frac{R}{r}\right)^{l+1} P_l^1(\cos\theta). \qquad (A.31)$$

On the right-hand side of Eq. (A.31) the variables are separated. Let us introduce a similar separation on the left-hand side:
$$A(r,\theta) = \sum_{l=1}^{\infty} R_l(r)\Theta_l(\theta). \qquad (A.32)$$

We will calculate the following expression
$$\nabla_\varphi^2 R_l(r)\Theta_l(\theta) = \frac{\Theta_l(\theta)}{r^2}\frac{\partial}{\partial r}\left(r^2 \frac{\partial R_l(r)}{\partial r}\right) + \frac{R_l(r)}{r^2 \sin\theta}\frac{\partial}{\partial \theta}\left(\sin\theta \frac{\partial \Theta_l(\theta)}{\partial \theta}\right) - \frac{R_l(r)\Theta_l(\theta)}{r^2 \sin^2\theta}. \qquad (A.33)$$

Let (internal problem
$$R_l(r) = \left(\frac{r}{R}\right)^{l+2}. \qquad (A.34)$$

Then Eq. (A.33) can be rewritten as
$$\nabla_\varphi^2 R_l(r)\Theta_l(\theta) = \frac{1}{R^2}\left(\frac{r}{R}\right)^l \left\{\frac{1}{\sin\theta}\frac{\partial}{\partial\theta}\left(\sin\theta\frac{\partial \Theta_l(\theta)}{\partial\theta}\right) - \frac{\Theta_l(\theta)}{\sin^2\theta} + (l+2)(l+3)\Theta_l(\theta)\right\}. \qquad (A.35)$$

It is obvious that to comply with Eq. (A.30) the relation must be satisfied
$$\frac{1}{\sin\theta}\frac{\partial}{\partial\theta}\left(\sin\theta\frac{\partial \Theta_l(\theta)}{\partial\theta}\right) - \frac{\Theta_l(\theta)}{\sin^2\theta} + (l+2)(l+3)\Theta_l(\theta) = -R^2 B_l P_l^1(\cos\theta). \qquad (A.36)$$

Introducing the definition $t = \cos\theta$, rewrite the left-hand side of Eq.(A.36) as
$$\frac{\partial}{\partial t}\left((1-t^2)\frac{\partial \Theta_l(\theta)}{\partial t}\right) - \frac{\Theta_l(\theta)}{1-t^2} + (l+2)(l+3)\Theta_l(\theta) =$$
$$= (1-t^2)\frac{\partial^2 \Theta_l(\theta)}{\partial t} - 2t\frac{\partial \Theta_l(\theta)}{\partial t} - \frac{\Theta_l(\theta)}{1-t^2} + (l+2)(l+3)\Theta_l(\theta) \qquad (A.37)$$

Let us compare expression (A.37) with the equation for the associated Legendre polynomials [20]:
$$(1-t^2)\frac{\partial^2 P_l^m(t)}{\partial t} - 2t\frac{\partial P_l^m(t)}{\partial t} + \left[l(l+1) - \frac{m^2}{1-t^2}\right]P_l^m(t) = 0. \qquad (A.38)$$

For $m=1$ we have
$$(1-t^2)\frac{\partial^2 P_l^1(t)}{\partial t} - 2t\frac{\partial P_l^1(t)}{\partial t} - \frac{P_l^1(t)}{1-t^2} + (l+2)(l+3)P_l^1(t) = \left[(l+2)(l+3) - l(l+1)\right]P_l^1(t) \qquad (A.39)$$

or

$$(1-t^2)\frac{\partial^2 P_l^1(t)}{\partial t} - 2t\frac{\partial P_l^1(t)}{\partial t} - \frac{P_l^1(t)}{1-t^2} + (l+2)(l+3)P_l^1(t) = (4l+6)P_l^1(t). \qquad (A.40)$$

By choosing

$$\Theta_l(\theta) = -\frac{R^2 A_l}{4l+6} P_l^1(\cos\theta) \qquad (A.41)$$

we will satisfy Eq. (A.36).

Substituting Eqs. (A.34) and (A.41) into Eq. (A.32), we obtain the desired solution for the vortex vector potential that determines the transverse velocity

$$A = -R^2 \sum_{l=1}^{\infty} \frac{A_l}{4l+6}\left(\frac{r}{R}\right)^{l+2} P_l^1(\cos\theta). \qquad (A.42)$$

For the external problem, instead of Eq. (A.34) we take

$$R_l(r) = \left(\frac{R}{r}\right)^{l-1}. \qquad (A.43)$$

Proceeding in a similar manner, we arrive at a relation for the vector potential of the form

$$A = R^2 \sum_{l=1}^{\infty} \frac{B_l}{4l-2}\left(\frac{R}{r}\right)^{l-1} P_l^1(\cos\theta). \qquad (A.44)$$

**A.3. Calculation of velocity**

The transverse velocity is calculated using Eqs. (A.26) and (A.27). In this case, making the replacement $t = \cos\theta$, we obtain

$$V_{\perp r} = -\frac{1}{r}\frac{\partial}{\partial t}[A(r,t)\sqrt{1-t^2}], \qquad (A.45)$$

$$V_{\perp \theta} = -\frac{1}{r}\frac{\partial}{\partial r}[rA(r,t)]. \qquad (A.46)$$

By definition, for the associated Legendre polynomials [20] we have:

$$P_l^m(t) = (-1)^m (1-t^2)^{\frac{m}{2}} \frac{d^m P_l(t)}{dt^m}. \qquad (A.47)$$

Using the recurrence relations [20], we obtain

$$\frac{\partial}{\partial t}\left(\sqrt{1-t^2} P_l^1(t)\right) = -\frac{P_l^1(t)t}{\sqrt{1-t^2}} + \sqrt{1-t^2}\frac{\partial}{\partial t} P_l^1(t) = l(l+1)P_l(t). \qquad (A.48)$$

Then, taking into account (A.42) and (A.48), we find

$$V_{\perp rl} = -l(l+1)R\frac{A_l}{4l+6}\left(\frac{r}{R}\right)^{l+1} P_l(t), \quad l=1,2,3\ldots \qquad (A.49)$$

Substituting (A.42) into (A.46), we obtain the expression for the polar component

$$V_{\perp \theta l} = -RA_l\frac{l+3}{4l+6}\left(\frac{r}{R}\right)^{l+1} P_l^1(t), \quad l=1,2,3\ldots \qquad (A.50)$$

Similarly, for the external problem, substituting Eq. (A.44) into Eq.(A.45), we obtain

$$V_{\perp rl} = -l(l+1)R \frac{B_l}{4l-2}\left(\frac{R}{r}\right)^l P_l(t). \qquad l=1,2,3\ldots \qquad (A.51)$$

Next, for the polar component of velocity, substituting (A.44) into (A.46), we find

$$V_{\perp \theta l} = RB_l \frac{l-2}{4l-2}\left(\frac{R}{r}\right)^l P_l^1(t). \qquad l=1,2,3\ldots \qquad (A.52)$$

Let us consider a solution for the longitudinal (irrotational) velocity (A.12), which includes the scalar potential $f$ satisfying the scalar Laplace equation (A.10), which in expanded form has the form:

$$\frac{1}{r^2}\frac{\partial}{\partial r}\left(r^2 \frac{\partial f}{\partial r}\right) + \frac{1}{r^2 \sin\theta}\frac{\partial}{\partial \theta}\left(\sin\theta \frac{\partial f}{\partial \theta}\right) = 0, \qquad (A.53)$$

and its solution for the internal problem, as is known, is written in the form [20]:

$$f(r,\theta) = R^2 \sum_{l=0}^{\infty} C_l \left(\frac{r}{R}\right)^l P_l. \qquad (54)$$

Here the term at $l=0$ is a constant and vanishes upon differentiation. For the radial component of the potential part of the velocity vector we have:

$$V_{\parallel r} = \frac{\partial}{\partial r} f(r,\theta) = R^2 \sum_{l=1}^{\infty} lC_l \frac{r^{l-1}}{R^l} P_l. \qquad (A.55)$$

For the polar component

$$V_{\parallel \theta} = \frac{1}{r}\frac{\partial}{\partial \theta} f(r,\theta) = -R^2 \sum_{l=1}^{\infty} C_l \frac{r^{l-1}}{R^l} \sin\theta \frac{dP_l(\cos\theta)}{d\cos\theta} \qquad (A.56)$$

or, using the definition of the associated Legendre polynomial (A.47),

$$V_{\parallel \theta} = R^2 \sum_{l=1}^{\infty} C_l \frac{r^{l-1}}{R^l} P_l^1. \qquad (A.57)$$

Similarly, for the external problem the solution is of the form

$$f(r,\theta) = R^2 \sum_{l=0}^{\infty} D_l \left(\frac{R}{r}\right)^{l+1} P_l. \qquad (A.58)$$

Then for the radial component of the potential part of the velocity vector we have:

$$V_{\parallel r} = \frac{\partial}{\partial r} f(r,\theta) = -R \sum_{l=0}^{\infty} (l+1)D_l \left(\frac{R}{r}\right)^{l+2} P_l. \qquad (A.59)$$

For the polar component

$$V_{\parallel \theta} = R \sum_{l=1}^{\infty} D_l \left(\frac{R}{r}\right)^{l+2} P_l^1(\cos\theta). \qquad (A.60)$$

Taking into account Eqs. (A.9), (A.49), (A.50), (A.59) and (A.60), the fluid flow velocity within the internal problem can be represented as the sum of the transverse and longitudinal components

$$V_r = \sum_{l=1}^{\infty} l(l+1)\left\{\frac{a_l}{4l+6}\left(\frac{r}{R}\right)^{l+1} + c_l\left(\frac{r}{R}\right)^{l-1}\right\} P_l, \qquad (A.61)$$

$$V_\theta = \sum_{l=1}^{\infty} \left\{ a_l \frac{l+3}{4l+6} \left(\frac{r}{R}\right)^{l+1} + c_l(l+1)\left(\frac{r}{R}\right)^{l-1} \right\} P_l^1. \qquad (A.62)$$

Here the coefficients have been redesignated to a more convenient form:

$$a_l = -RA_l, \qquad c_l = \frac{RC_l}{l+1}. \qquad (A.63)$$

Similarly, for the external problem, using Eqs. (A.88), (A.89) and (A.89):

$$V_r = -\sum_{l=1}^{\infty} l(l+1)\left\{ \frac{b_l}{4l-2}\left(\frac{R}{r}\right)^{l} + d_l\left(\frac{R}{r}\right)^{l+2} \right\} P_l(t) + d_0\left(\frac{R}{r}\right)^2. \qquad (A.64)$$

$$V_\theta = \sum_{l=1}^{\infty} \left\{ b_l \frac{l-2}{4l-2}\left(\frac{R}{r}\right)^{l} + d_l l\left(\frac{R}{r}\right)^{l+2} \right\} P_l^1(t). \qquad (A.65)$$

Here the coefficients have been redesignated:

$$b_l = RB_l, \qquad d_l = \frac{RD_l}{l} \text{ при } l=1,2,3\ldots: \qquad d_0 = -RD_0 \qquad (A.66)$$

**A.4. Calculation of pressure**

The pressure satisfies the Laplace equation (A.6), which in the case under consideration has the form:

$$\frac{1}{r^2}\frac{\partial}{\partial r}\left(r^2 \frac{\partial p}{\partial r}\right) + \frac{1}{r^2 \sin\theta}\frac{\partial}{\partial \theta}\left(\sin\theta \frac{\partial p}{\partial \theta}\right) = 0. \qquad (A.67)$$

Its solution is [20]

$$p(r,\theta) = \sum_{l=0}^{\infty} f_l \left(\frac{r}{R}\right)^l P_l(\cos\theta). \qquad (A.68)$$

It was shown earlier (Eq. (A.17)) that the pressure gradient in the Stokes equation is determined only by the transverse (vortex) component of the velocity. Therefore, the coefficients $f_l$ in Eq. (A.68) are proportional to the coefficients $a_l$ in Eqs. (A.61) and (A.62) that characterize the transverse (vortex) part of the velocity.

The Stokes equation (A.17) for an axisymmetric problem has the form

$$\nabla^2 V_{\perp r} - \frac{2V_{\perp r}}{r^2} - \frac{2}{r^2 \sin\theta}\frac{\partial}{\partial\theta}(V_{\perp\theta}\sin\theta) = \eta^{-1}\frac{\partial}{\partial r}p, \qquad (A.69)$$

$$\nabla^2 V_{\perp\theta} - \frac{V_{\perp\theta}}{r^2 \sin^2\theta} + \frac{2}{r^2}\frac{\partial}{\partial\theta}V_{\perp r} = \eta^{-1}\frac{1}{r}\frac{\partial}{\partial\theta}p, \qquad (A.70)$$

where, as follows from Eqs. (A.61) and (A.62), the projections of the transverse velocity are determined by the expressions

$$V_{\perp r} = \sum_{l=1}^{\infty} l(l+1)\frac{a_l}{4l+6}\left(\frac{r}{R}\right)^{l+1} P_l, \qquad (A.71)$$

$$V_{\perp\theta} = \sum_{l=1}^{\infty} a_l \frac{l+3}{4l+6}\left(\frac{r}{R}\right)^{l+1} P_l^1. \qquad (A.72)$$

For the pressure gradient (A.68) we obtain:
radial component

$$\frac{\partial p}{\partial r} = \sum_{l=1}^{\infty} \frac{lf_l}{R}\left(\frac{r}{R}\right)^{l-1} P_l(\cos\theta), \tag{A.73}$$

polar component

$$\frac{1}{r}\frac{\partial p}{\partial \theta} = \sum_{k=1}^{\infty} \frac{f_l}{R}\left(\frac{r}{R}\right)^{l-1} P_l^1(\cos\theta). \tag{A.74}$$

Substituting Eqs. (A.71)-(A.74) into Eqs. (A.69) and (A.70) allows us to express the coefficients $f_l$ in terms of $a_l$. Making transformations, using known recurrence relations and the Legendre equation [20], we arrive at the following coefficient relation

$$f_l = \eta\frac{(l+1)}{R}a_l. \tag{A.75}$$

Hence

$$p(r,\theta) = \frac{\eta}{R}\sum_{l=0}^{\infty}(l+1)a_l\left(\frac{r}{R}\right)^l P_l(\cos\theta). \tag{A.76}$$

Expression (A.76) together with Eqs. (A.61) and (A.62) give the desired general solution of the internal axisymmetric problem in spherical coordinates.

The external problem is considered similarly. The transverse velocity components according to Eqs. (A.64) and (A.65) have the form

$$V_{\perp r} = -\sum_{l=1}^{\infty} l(l+1)\frac{b_l}{4l-2}\left(\frac{R}{r}\right)^l P_l, \tag{A.77}$$

$$V_{\perp\theta} = \sum_{l=1}^{\infty} b_l \frac{l-2}{4l-2}\left(\frac{R}{r}\right)^l P_l^1. \tag{A.78}$$

For the pressure in this case we obtain the following solution

$$p(r,\theta) = -\frac{\eta}{R}\sum_{l=1}^{\infty} lb_l\left(\frac{R}{r}\right)^{l+1} P_l(\cos\theta). \tag{A.79}$$

Eqs. (A.79), (A.64) and (A.65) are the general solution of the external axisymmetric problem in spherical coordinates.

### A.5. Calculation of stream function

For the velocity components expressed through the stream function, we have [11-13]:

$$V_r = \frac{1}{r^2 \sin\theta}\frac{\partial \Psi}{\partial \theta}. \tag{A.80}$$

$$V_\theta = -\frac{1}{r\sin\theta}\frac{\partial \Psi}{\partial r}. \tag{A.81}$$

In the approach being developed, the stream function is a secondary, auxiliary quantity. It is calculated here a posteriori. It is useful to know it in order to plot streamlines. To obtain the stream function corresponding to the transverse part of the velocity, it is necessary to use Eqs. (A.26)-(A.27) that establish the connection between the transverse velocity and the φ-component of the vector potential. Comparing with Eqs. (A.80)-(A.81), we obtain:

$$\Psi_\perp(r,\theta) = rA(r,\theta)\sin\theta. \tag{A.82}$$

The part of the function corresponding to the longitudinal part of the velocity can be obtained from Eqs. (A.80)-(A.81) taking into account Eq. (A.12):

$$\frac{\partial f}{\partial r} = \frac{1}{r^2 \sin\theta} \frac{\partial \Psi_\parallel}{\partial \theta}, \tag{A.83}$$

$$\frac{\partial f}{\partial \theta} = -\frac{1}{\sin\theta} \frac{\partial \Psi_\parallel}{\partial r}. \tag{A.84}$$

The full stream function is found as the sum

$$\Psi = \Psi_\perp + \Psi_\parallel. \tag{A.85}$$

Carrying out the indicated procedure, taking into account Eqs. (A.61)-(A.62), for the internal problem we obtain a stream function of the form

$$\Psi(r,\theta) = -R^2 \sin\theta \sum_{l=1}^{\infty} \left\{ \frac{a_l}{4l+6} \left(\frac{r}{R}\right)^{l+3} + c_l \left(\frac{r}{R}\right)^{l+1} \right\} P_l^1(\cos\theta). \tag{A.86}$$

Taking into account Eqs. (A.64)-(A.65), for the external problem we obtain the following expression

$$\Psi(r,\theta) = R^2 \sin\theta \sum_{l=1}^{\infty} \left\{ \frac{b_l}{4l-2} \left(\frac{R}{r}\right)^{l-2} + d_l \left(\frac{R}{r}\right)^l \right\} P_l^1(\cos\theta) - d_0 R^2 \cos\theta. \tag{A.87}$$

This completes the derivation of the main relationships. The practically important equations that were obtained here are placed in Table 1.

**AUTHOR DECLARATIONS**

**Conflict of Interest**

The authors have no conflicts to disclose.

**DATA AVAILABILITY**

The data that support the findings of this study are available from the corresponding authors upon reasonable request.